\documentclass[traditabstract]{aa} 
\usepackage{graphicx}
\usepackage{txfonts}
\usepackage{natbib,twoopt}
\usepackage{textcomp}
\usepackage{longtable}
\usepackage{booktabs}
\usepackage{lscape}
\usepackage{xspace}
\usepackage{tabularx}
\begin{document}
\title{Eclipsing high-mass binaries}
\subtitle{I. Light curves and system parameters for CPD\,$-$\,51\degr\,8946, PISMIS\,24-1
and HD\,319702}

	\titlerunning{Eclipsing high-mass binaries}

	\author{
          A. Barr Dom\'{i}nguez
          \inst{1}
          \and
          R. Chini
          \inst{1,2}
          \and
          F. Pozo Nu\~nez
          \inst{1}		
          \and
          M. Haas
          \inst{1}
          \and
          M. Hackstein
          \inst{1}
          \and
          H. Drass
          \inst{1}
          R. Lemke
          \inst{1}		
          \and
          M. Murphy
          \inst{2}		
	}
	\institute{
          Astronomisches Institut, Ruhr--Universit\"at Bochum,
	  Universit\"atsstra{\ss}e 150, 44801 Bochum, Germany
	  \and
          Instituto de Astronomia, Universidad Cat\'{o}lica del
          Norte, Avenida Angamos 0610, Casilla
          1280 Antofagasta, Chile
        }

	\authorrunning{A. Barr Dom\'{i}nguez et al.}

	\date{Received ; accepted}

	\abstract{ We present first results of a comprehensive photometric O-star survey performed with a robotic twin refractor at the Universit\"atssternwarte Bochum located near Cerro Armazones in Chile. For three high-mass stars, namely Pismis\,24-1, CPD\,$-51\degr\,8946$ and HD\,319702, we determined the period through the Lafler-Kinman algorithm and model the light curves within the framework of the Roche geometry.
For Pismis\,24-1, a previously known eclipsing binary, we provide first light curves and determined a photometric period of 2.36 days together with an orbital inclination of $61.8^\circ$. The best-fitting model solution to the light curves suggest a detached configuration. With a primary temperature of $T_1 = 42520$\,K we obtain the temperature of the secondary component as $T_2 = 41500$\,K.
CPD\,$-$\,51\degr\,8946 is another known eclipsing binary for which we present a revised photometric period of 1.96 days with an orbital inclination of $58.4^\circ$. The system has likely a semi-detached configuration and a mass ratio $q = M_1/M_2 = 2.8$. If we adopt a primary temperature of $T_1 = 34550$\,K we obtain $T_2 = 21500$\,K for the secondary component.
HD\,319702 is a newly discovered eclipsing binary member of the young open cluster NGC\,6334. The system shows well-defined eclipses favouring a detached configuration with a period of 2.0 days and an orbital inclination of $67.5^\circ$. Combining our photometric result with the primary spectral type O8 III(f) ($T_1 = 34000$\,K) we derive a temperature of $T_2 = 25200$\,K for the secondary component.
        }

		\maketitle

\section{Introduction}

The fundamental quantity of a star is its mass because it determines the energy production, the evolution and the final state; certainly there are other parameters like chemical composition and angular momentum that influence stellar evolution. At first glance, the easiest way to estimate stellar masses is by means of the mass-luminosity relation $L \propto M^\alpha$ which is well established by both theory and observation. However, the exponent $\alpha$ itself is a function of mass: while $\alpha \sim 4$ for solar-type stars it becomes smaller for stars of lower ($< 0.5\,M_\odot$) and higher ($> 10\,M_\odot$) mass. In practice there are many other parameters that influence the mass-luminosity relation. All stars increase their luminosity with age -- on the other hand high-mass stars lose a considerable fraction of their mass during evolution. Observationally, the determination of the bolometric luminosity ($L_{\rm bol}$) becomes more difficult for O- and early B-type stars because due to their high temperature most light is radiated in the $UV$. Likewise, distance, reddening and effective temperature ($T_{\rm eff}$) have to be known in order to calculate the absolute visual magnitude $M_V$ and to convert it into $L_{\rm bol}$. For high-mass stars one can circumvent the mass-luminosity relation and determine $T_{\rm eff}$ and $L_{\rm bol}$ by modeling the optical spectra and by placing the star in the Hertzsprung-Russel diagram (HRD). Eventually evolutionary models yield the stellar mass which is obviously based on many assumptions and whose error is dominated by the uncertainties of theoretical models.

In contrast, binary stars allow the determination of fundamental stellar parameters through a combination of photometry, astrometry, and spectroscopy and by applying basic physical laws. In case of an eclipsing binary (hereafter EB), radii, distance and, in favorable cases, $T_{\rm eff}$ also may be determined from a combined analysis of light curves and radial velocity (hereafter $RV$) curves.

There is growing evidence that most high-mass stars occur as binaries and/or multiple systems (e.g., \citealt{1999NewA....4..531P}; \citealt{2009AJ....137.3358M}; \citealt{2011IAUS..272..474S}; \citealt{2012MNRAS.424.1925C}). Because properties such as binary fraction, period distribution, and mass-ratio distribution provide important constraints on models of star formation and dynamical evolution, precise knowledge of multiplicity characteristics and how they change within this exceptional mass region are important to understanding the formation of high-mass stars. The currently discussed high-mass star formation scenarios involve: i) the gravitational collapse of isolated massive cores and accretion disk (e.g., \citealt{1991ASPC...13...73M}; \citealt{2002Sci...295...76W}; \citealt{2002ApJ...569..846Y}; \citealt{2005astro.ph.10412K}; \citealt{2006astro.ph..7429K}), ii) competitive accretion in a clustered environment (\citealt{2003MNRAS.343..413B}; \citealt{2006MNRAS.370..488B}; \citealt{2008ASPC..390...26B}; \citealt{2008ASPC..387..208C}), and iii) stellar collisions in very dense clusters (\citealt{1998MNRAS.298...93B}; \citealt{2000prpl.conf..327S}; \citealt{2009Ap&SS.324..271V}).

Once a star has evolved to where it fills its Roche lobe, interaction with its companion through the transfer of mass is inevitable. The evolution of interacting binaries follows a significantly different path from that of single stars because of the mass flow outwards from one component, the mass gain by its companion and the mass loss from the entire system. The evolution of massive stars in binary systems may lead to exotic phenomena such as stellar mergers, X-ray binaries or gamma-ray bursts. As recently claimed by \cite{2012Sci...337..444S} more than 70\% of all high-mass stars will exchange mass with a companion, leading to a binary merger in one-third of the cases.

While great progress has been made so far in a statistical sense, the system parameters like period ($P$), eccentricity ($e$) and mass for individual high-mass binaries are less known. To date, substantial parts of the observational material for binaries and multiple systems among high-mass stars comes from high-resolution spectroscopic monitoring campaigns performed for clusters e.g.
\object{Trumpler\,14} and
\object{Trumpler\,16} (\citealt{2001MNRAS.326.1149R}),
\object{Cr\,228} (\citealt{1990ApJS...72..323L}),
\object{IC\,2944} (\citealt{2011MNRAS.416..817S}),
\object{IC\,1805} (\citealt{2006A&A...456.1121D}),
\object{NGC\,6231} (\citealt{2008MNRAS.386..447S}),
\object{NGC\,6611} (\citealt{2009A&A...501..291S}),
\object{NGC\,2244} (\citealt{2009A&A...502..937M}),
\object{NGC\,6334} and
\object{NGC\,6357} (\citealt{2012A&A...538A.142R})
or OB associations e.g.
\object{Monoceros\,OB2} (\citealt{2009A&A...502..937M})
and
\object{Cyg\,OB2} (\citealt{2009A&A...502..937M}).

Spectroscopic orbital parameters of individual binary systems have been reported for e.g.
HD\,166734 (O+O) (\citealt{1980ApJ...238..184C}),
V382 Cyg (O+O),
V448 Cyg (O+O),
XZ Cep (O+O) (\citealt{1997MNRAS.285..277H}),
HD\,93403 (O+O) (\citealt{2000A&A...360.1003R}),
HD\,149404 (O+O) (\citealt{2001A&A...368..212R}),
HD\,152248 (O+O) (\citealt{2001A&A...370..121S}),
HD\,101131 (O+O) (\citealt{2002ApJ...574..957G}),
HD\,48099 (O+O) (\citealt{2010ApJ...708.1537M}),
HD\,152219 (O+B)(\citealt{2006MNRAS.371...67S}),
HD\,115071 (O+B) (\citealt{2002ApJ...575.1050P}).

Photometric orbital parameters were reported e.g. by
\cite{1988MNRAS.235..797D},
\cite{1992MNRAS.254..404B},
\cite{2003AJ....126.2988T}, and
\cite{2006MNRAS.371...67S}.

There are only a few dozen EBs known among the O-type stars in the Galaxy, e.g.
HD\,167971 (O+O) (\citealt{1988MNRAS.235..797D}),
V1182\,Aqu (O+O) (\citealt{2005ApJS..161..171M}),
FO15       (O+O) (\citealt{2006MNRAS.367.1450N}), and
SZ\,Cam    (O+B) (\citealt{1998A&A...332..909L}).
Further O-type EBs are desirable to increase the statistics on this exotic class of objects.

Currently, we are performing a photometric monitoring survey with the aim to detect all eclipsing O-type binaries in the southern hemisphere from two brightness limited samples. Our targets comprise a complete sample of about 250 O-type stars ($V < 8$) taken from the Galactic O-Star Catalogue V.2.0 (\citealt{2008RMxAC..33...56S}). Preliminary results indicate variability for about 24\% of the objects (\citealt{2013CEAB...37..295C}). A second sample of fainter O stars comes from the Bochum Galactic disk Survey (\citealt{2012AN....333..706H}) where we are monitoring all stars with $10 < R,I < 15$ in a strip of $\Delta b = \pm 3^\circ$ along the southern galactic plane.

In this paper we present a detailed photometric multi-epoch study of three selected O-type stars, \object{HD319702}, \object{CPD\,$-$\,51\degr\,8946} and \object{PISMIS\,24-1}. We show the first light curve for Pismis\,24-1, re-analyze the orbital period for CPD\,$-$\,51\degr\,8946 and finally present the orbital parameters for a new high-mass EB system HD\,319702.

\section{Data}

\subsection{Observations}

The photometric observations were conducted between May and October 2011 using the robotic 15\,cm VYSOS-6 telescope of the Universit\"atssternwarte Bochum, located near Cerro Armazones, the future location of the ESO Extreme Large Telescope (ELT) in Chile\footnote{http://www.astro.ruhr-uni-bochum.de/astro/oca/}; for instrumental details see \cite{2012AN....333..706H}. The images were obtained simultaneously through Sloan $r$ and $i$ filters at 6230\,\AA\ and 7616\,\AA; each star was observed typically for about 40 epochs.

\subsection{Reduction}

The images were processed by standard IRAF\footnote{IRAF is distributed by the National Optical Astronomy Observatory, which is operated by the Association of Universities for Research in Astronomy (AURA) under cooperative agreement with the National Science Foundation.} routines for image reduction, including bias, dark current and flatfield correction. Additionally, astrometry and astrometric distortion were calculated and properly corrected using SCAMP (\citealt{2006ASPC..351..112B}) in combination with SExtractor (\citealt{1996A&AS..117..393B}). To improve the image quality further we resampled the frames of an original pixel size of 2.4\arcsec\, to a new pixel size of 0.8\arcsec\, using the routine SWARP (\citealt{2002ASPC..281..228B}). More information about the data reduction can be found in \cite{2012AN....333..706H}. Photometry was performed using an aperture radius of 4\arcsec\, maximizing the signal-to-noise ratio ($S/N$) and delivering the lowest absolute scatter for the fluxes.

The light curves in normalized flux units are calculated relative to nearby non-variable reference stars located in the same field. The absolute photometric calibration was obtained using the fluxes of about 20 standard stars from \cite{2009AJ....137.4186L}. The standard star fields have been observed during the same nights as the science targets. The photometry was corrected for airmass by using the extinction curve for the nearby site Cerro Paranal derived by \cite{2011A&A...527A..91P}. The photometric errors are typically of the order of 0.05\,mag at both wavelengths.

\subsection{Light Curve analysis}

The photometric light curves were analysed assuming a standard Roche geometry based on the Wilson-Devinney code (\citealt{1971ApJ...166..605W}; \citealt{1979ApJ...234.1054W}, \citealt{1990ApJ...356..613W}). Because Kurucz model atmospheres assume local thermodynamic equilibrium (LTE) for calculating the emergent fluxes from a star this approximation is insufficient in estimating the resultant light curves of high-mass stars, where non-LTE effects alter the radiation field significantly from that of LTE. This is even more prominent in evolved stars, like Pismis\,24-1, which is one of the objects investigated in the present study. For a more thorough
analysis of these systems, after follow-up spectroscopy has been obtained, we will use a code that allows for input of model atmospheres that do not make the LTE assumption, such as the ELC code (\citealt{2000A&A...364..265O}).

For each model, the period $P$ of the system and $T_{\rm eff}$ of one component were fixed. In the following we refer to the temperatures of the primary and secondary component as $T_1$ and $T_2$, respectively. Periods were determined independently through the Lafler-Kinman algorithm (\citealt{1965ApJS...11..216L}) which was subsequently generalized and called Phase Dispersion Minimization (PDM) by \cite{1978ApJ...224..953S}. This method selects the period that yields the lowest dispersion of the phase light curve. The advantage of this algorithm compared to others (e.g. Analysis of Variance by \cite{1989MNRAS.241..153S}, \cite{1999ApJ...516..315S}; Lomb-Scargle periodogram by \cite{1982ApJ...263..835S}) is that it works also reliably in cases where
there are only a moderate number of data points ($\sim 50$).

Effective temperatures were adopted according to the spectral types of the stars. For cases where the mass ratio ($q = M_1/M_2)$ of the stars was known, $T_{\rm eff}$ was also fixed accordingly. The best model was determined through several fits to the light curves until the minimum of the $\chi^2$ was reached for each of the following free parameters adopted: the effective temperature of the secondary star, the Roche lobe filling factors and the inclination $i$. Because our light curves do not reveal any signs for spots, any further physical analysis was excluded.

A standard gravity-darkening law ($T_{\rm eff} \sim g^{\beta}$) with coefficients $\beta_{1} = \beta_{2} = 0.25$ (\citealt{1924MNRAS..84..702V}; \citealt{2000A&A...359..289C}; \citealt{2003A&A...402..667D}) together with bolometric albedos $A_1 = A_2 = 1.0$ was assumed for early-type stars with radiative envelopes and hydrostatic equilibrium. We used a non-linear
square-root limb-darkening law obtained at optical wavelengths (\citealt{1992A&A...259..227D}) with limb darkening coefficients interpolated from tables of \cite{1993AJ....106.2096V} at the given bandpass.

\section{Results}

In the following, we introduce the three multiple high-mass systems and describe their properties as obtained from our light curve analysis.

\begin{table}[h!]
\caption{\object{Pismis\,24-1} orbital solution and system parameters. The reference time, $T_{0}$, refers to the time of the
primary eclipse.}
\label{table2}
\begin{tabular}{lllll}
\hline
\hline
\noalign{\smallskip}
Parameters & Sloan~$r$ & Sloan~$i$ &   &  \\
\noalign{\smallskip}
\hline
\noalign{\smallskip}
$P$ [d] & $2.358\pm0.007$ & $2.359\pm0.009$ & & \\
$T_{0}$(HJD-2400000) & $55697.2788$ & $55697.2788$ & &  \\
$q (M_1/M_2)$ & $1.0$ & $1.0$ & & \\
$i$ & $61\fdg7 \pm 1\fdg5$ & $61\fdg8 \pm 1\fdg5$ & & \\
$e$ & $0.0$ & $0.0$ & & \\
Roche lobe coefficient(1) & $0.791\pm0.006$ & $0.793\pm0.006$ & & \\
Roche lobe coefficient(2) & $0.810\pm0.007$ & $0.821\pm0.007$ & & \\
$T_{1}$ (K) & $42\,520$ & $42\,520$ & & \\
$T_{2}$ (K) & $41\,500 \pm 1260$ & $41\,500 \pm 1267$ & & \\
$\Delta D_{p}$              & $8.0\%$ & $7.0\%$ & & \\
$\Delta D_{s}$              & $8.0\%$ & $7.0\%$ & & \\
\hline
\end{tabular}
\end{table}

\subsection{{\bf Pismis\,24-1}}

\object{Pismis\,24-1} is a triple system that belongs to the young open cluster \object{Pismis\,24} which resides within NGC\,6357 in the Sagittarius spiral arm at a distance of 1.7\,kpc (\citealt{2012A&A...539A.119F}). This well-studied region is known as a rich reservoir of young stars, containing several OB-type stars (\citealt{2001AJ....121.1050M}) and several shell-like H\,{\sc ii} regions (\citealt{2012A&A...538A.142R}).

As part of a multiplicity study at the high-mass end of the IMF, Pismis\,24-1 (HDE\,319718\,A) has been investigated in depth by \cite{2007ApJ...660.1480M} by means of high-resolution images from the Hubble Space Telescope (HST). These observations resolved the system into two components -- \object{Pismis\,24-1\,SW} and \object{Pismis\,24-1\,NE} separated by 0.36\arcsec (see Fig.~1 of \citealt{2007ApJ...660.1480M}). Both components have similar absolute visual magnitudes ($M_{V} = -6.28$ for Pismis\,24-1\,SW and $M_{V} = -6.41$ for Pismis\,24-1\,NE) and similar intrinsic colors.

High-resolution spectroscopy (\citealt{2007ApJ...660.1480M}) yielded a spectral type O4\,III(f+) for \object{Pismis\,24-1\,SW} and confirmed the spectral type O3.5\,If* for \object{Pismis\,24-1\,NE} as determined previously by \cite{2002AJ....124..507W}. Radial velocity variations from $+20$ to $-90$\,km\,s$^{-1}$ for the absorption lines measured in eight spectra confirmed the variations reported previously by \cite{1984A&A...140...24L} and suggest that \object{Pismis\,24-1\,NE} is an unresolved spectroscopic binary. The estimated masses for \object{Pismis\,24-1\,SW} and \object{Pismis\,24-1\,NE} are $96 \pm 10\,M_\odot$ and $97 \pm 10\,M_\odot$ (\citealt{2007ApJ...660.1480M}).

Phil Massey and collaborators have detected optical variability in the unresolved Pismis\,24-1\,NE+SW system with a peak-to-peak amplitude of 0.07\,mag and a period of 2.36088 days\footnote{Private communication between \citealt{2007ApJ...660.1480M} and Phil Massey (2006).}. However, no light curve has been published to date. As a consequence, Pismis\,24-1 is an excellent target to provide a decent light curve for this system and to test our photometric analysis.

\begin{figure}[h!]
  \centering
  \includegraphics[angle=0,width=\columnwidth]{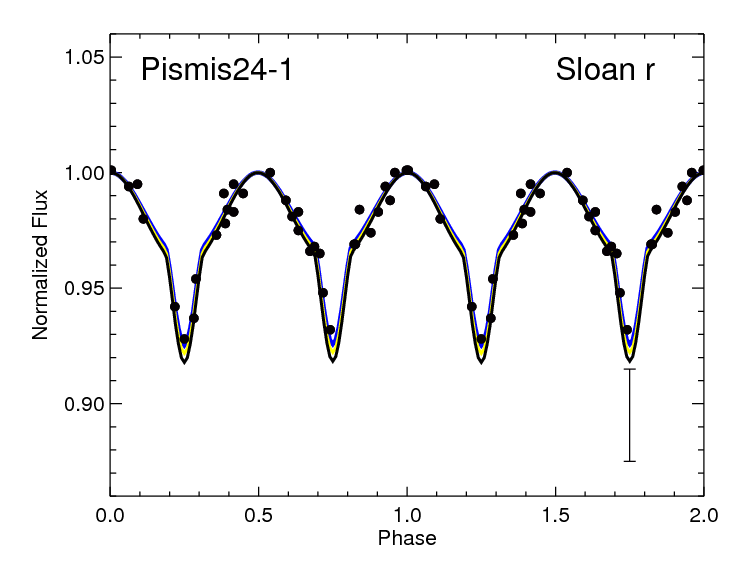}
  \includegraphics[angle=0,width=\columnwidth]{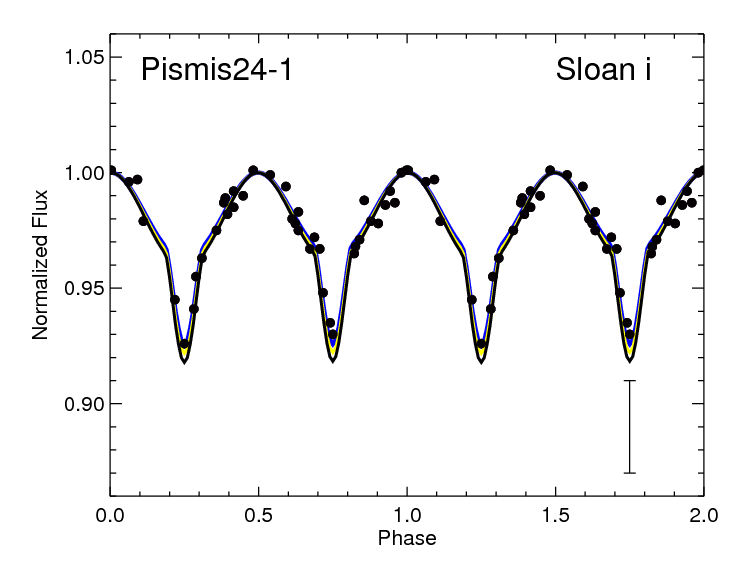}
  \includegraphics[angle=270,width=6cm]{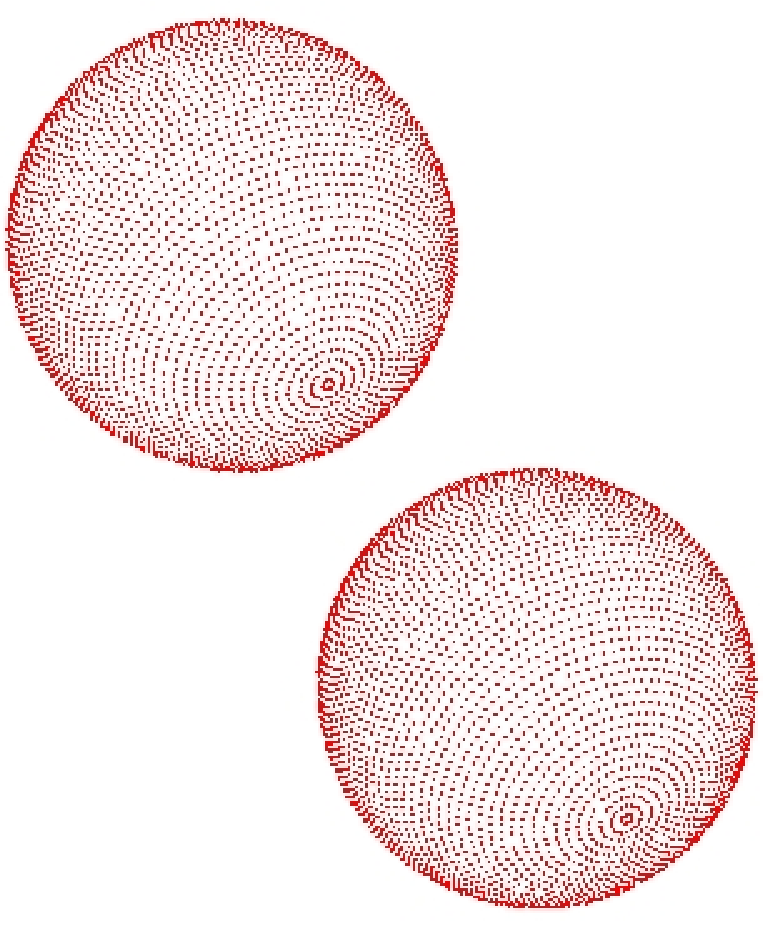}
  \caption{Observed light curves of Pismis\,24-1 with the best-fitting model obtained at $l_{3}=0.15$ (yellow solid line) in the Sloan~$r$ (top) and Sloan~$i$ (middle) bands, folded onto the orbital period of 2.36 days. The error bar in the bottom of the plot represents the average measurement uncertainty for the data set.
The blue and black solid lines correspond to the solutions obtained at $l_{3}=0.1$ and $l_{3}=0.2$ respectively. Bottom: 3D view of the system Pismis\,24-1 at an orbital phase of 0.12.}
  \label{Pismis24-1}
\end{figure}

The PDM analysis of our 32 data points yields a period of $P = 2.36$ days which is in excellent agreement with the result reported by \cite{2007ApJ...660.1480M}.

As already noted in Sect. 2, each final (reduced) image has a resolution of 0.8\arcsec which does not allow us to resolve both Pismis\,24-1\,NE+SW separately. In consequence, we expect that some amount of third light contribution could alter the properties of the system, reducing the depths of the eclipses resulting in temperature and inclination being underestimated. In order to estimate the third light contribution, we performed the analysis adopting $l_{3}$ as a free parameter. The analysis yields reasonable solutions for $0.05 \le l_{3} \le 0.20$ with the best-fitting model at $l_{3}=0.15$. An acceptable solution can be found without considering the third light component ($l_{3}=0$), however, the uncertainty in the inclination and the temperature is large. Fig.~\ref{Pismis24-1} displays the observed Sloan~$r$ and Sloan~$i$ light curves together with the best-fitting model solution folded onto the orbital period. The light curves in both filters show nearly the same amplitude variations ($\sim 7\%$) and both partial eclipses with the same duration of 0.174 in phase. The primary and secondary minimum eclipses, which are separated symmetrically by 0.5 in phase, show similar depths $\Delta D_{p} = \Delta D_{p,min} - \Delta D_{max} = 8\%$ and $\Delta D_{s} = \Delta D_{s,min} - \Delta D_{max} = 7\%$, where $p$ and $s$ refer to the primary and secondary minimum, and $max$ denotes to the global maximum of the light curve (\citealt{2003AJ....125.2173S}).

We fixed $T_1 = 42000\,K$ according to the spectral type O3.5\,If* reported by \cite{2007ApJ...660.1480M} and used the observational $T_{eff}$ calibration by \cite{2005A&A...436.1049M} (his Table 6). Additionally, we assumed equal masses for both components ($q = 1.0$) as suggested by the spectroscopy (\citealt{2007ApJ...660.1480M}). The best-fitting model favours a detached configuration with an inclination $i = 61.8^\circ$ and requires that both components fill up their Roche lobes at about 79\% and 81\%, respectively. A 3D graphic presentation of the Roche model of the system obtained at a orbital phase $\phi = 0.12$ is shown in Fig.~\ref{Pismis24-1}. Our calculations yield an effective temperature  of $T_2 = 41500\,K$ for the secondary star which is consistent with its spectral type designation. The spectral type O3.5\,If* corresponds to the combined light of the two stars; hence spectroscopic follow-up is required to classify the individual components. Likewise, the similarity of the depths in the primary and secondary minimum agrees well with the derived temperature. The observed symmetrical separation of the primary and secondary minimum eclipses suggests that the system has a circular orbit compatible with the current spectroscopic knowledge of the system. In consequence, we assumed zero eccentricity and synchronous rotation for which the model provides the best fit to the system. A summary of the relevant parameters and the best-fitting values are listed in Table~\ref{table2}.

\begin{table}
\caption{\object{CPD\,$-$\,51\degr\,8946} orbital solution and system parameters.The reference time, $T_{0}$, refers to the time of the primary eclipse.}
\label{table3}
\begin{tabular}{lllll}
\hline
\hline
\noalign{\smallskip}
Parameters & Sloan~$r$ & Sloan~$i$ &   &  \\
\noalign{\smallskip}
\hline
\noalign{\smallskip}
$P$ [d] & $1.959\pm0.009$ & $1.961\pm0.009$ & & \\
$T_{0}$(HJD-2400000) & $55699.2733$ & $55699.2733$ & & \\
$q (M_1/M_2)$ & $2.8\pm0.8$ & $2.8\pm0.8$ & & \\
$i$ & $58\fdg41 \pm 1\fdg2$ & $58\fdg41 \pm 1\fdg2$ & & \\
$e$ & $0.0$ & $0.0$ & & \\
Roche lobe coefficient(1) & $0.950\pm0.012$ & $0.950\pm0.012$ & & \\
Roche lobe coefficient(2) & $0.990\pm0.017$ & $0.990\pm0.017$ & & \\
$T_{1}$ (K) & $34\,550$ & $34\,550$ & & \\
$T_{2}$ (K) & $21\,500 \pm 1500$ & $21\,500 \pm 1500$ & & \\
$\Delta D_{p}$              & $16.0\%$ & $10.0\%$ & & \\
$\Delta D_{s}$              & $16.0\%$ & $10.0\%$ & & \\
\hline
\end{tabular}
\end{table}

\subsection{\bf CPD\,$-$\,51\degr\,8946}

\begin{figure}[h!]
  \centering
  \includegraphics[angle=0,width=\columnwidth]{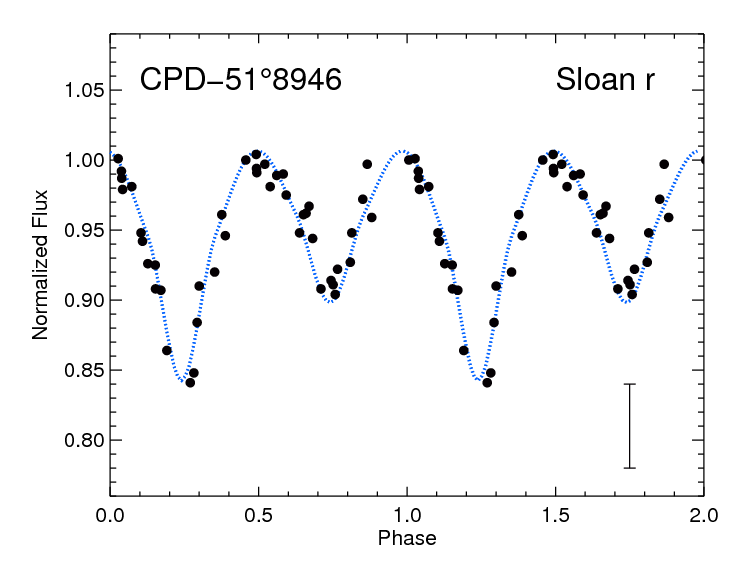}
  \includegraphics[angle=0,width=\columnwidth]{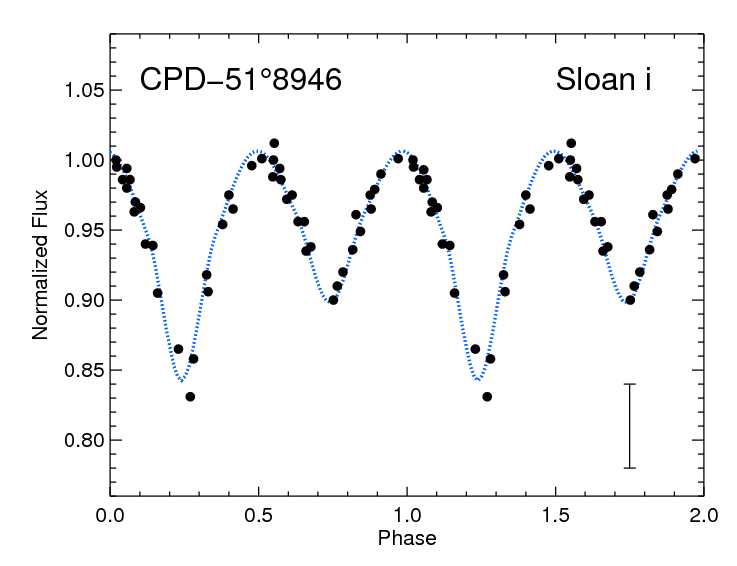}
  \includegraphics[angle=0,width=\columnwidth]{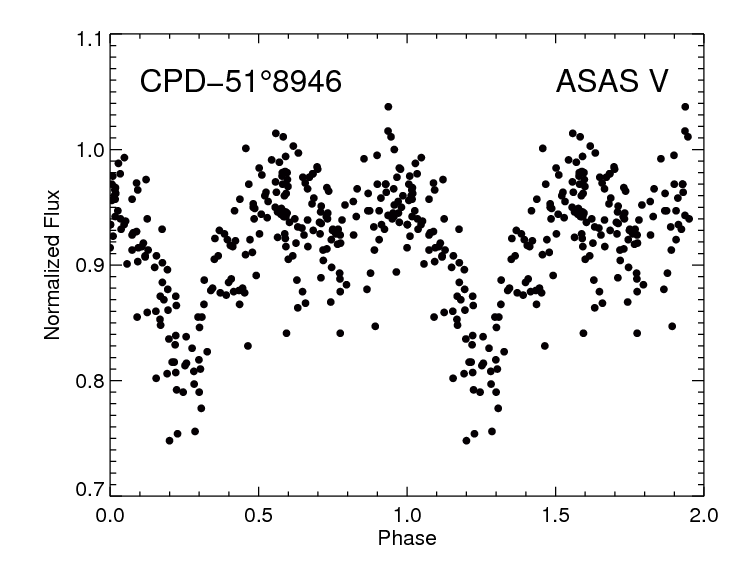}
  \caption{Observed (dots) and simulated (solid line) light curves in the r-Sloan (top) and i-Sloan (middle) bands, folded onto the orbital period ($1.96$ days) obtained from the PDM analysis. The error bar in the bottom of the plot represents the average measurement uncertainty for the data set. Bottom: The normalized V-band light curve obtained from the ASAS III catalogue folded with a period of $1.96$ days.
  }
  \label{CPD518946}
\end{figure}

\object{CPD\,$-51^\circ\,8946$} was classified as an OB-type star (\citealt{1964MeLu2.141....1L}) and is part of a sample of
about 700 OB stars observed by \cite{2001A&A...369..527V} at the former Leiden Southern Station at Hartebeespoortdam, South
Africa. The eclipsing nature of \object{CPD\,$-51^\circ\,8946$} was established by \cite{2004AcA....54..153P} using the data
obtained from the ASAS III (All Sky Automated Survey) catalogue of variable stars. The authors determined a photometric period of
3.92 days and suggest an eclipsing detached binary configuration.

\begin{table}
\caption{\object{HD\,319702} orbital solution and system parameters. The reference time, $T_{0}$, refers to the time of the primary
eclipse.}
\label{table4}
\begin{tabular}{lllll}
\hline
\hline
\noalign{\smallskip}
Parameters & Sloan~$r$ & Sloan~$i$ &   &  \\
\noalign{\smallskip}
\hline
\noalign{\smallskip}
$P$ [d] & $2.005\pm0.036$ & $2.006\pm0.036$ & & \\
$T_{0}$(HJD-2400000) & $55688.3448$ & $55688.3448$ & &  \\
$q (M_1/M_2)$ & $1.0$ & $1.0$ & & \\
$i$ & $67\fdg52 \pm 1\fdg1$ & $67\fdg52 \pm 1\fdg1$ & & \\
$e$ & $0.0$ & $0.0$ & & \\
Roche lobe coefficient(1) & $0.710\pm0.01$ & $0.710\pm0.01$ & & \\
Roche lobe coefficient(2) & $0.710\pm0.01$ & $0.710\pm0.01$ & & \\
$T_{1}$ (K) & $34\,000$ & $34\,000$ & & \\
$T_{2}$ (K) & $25\,200 \pm 1340$ & $25\,200 \pm 1340$ & & \\
$\Delta D_{p}$            & $11.0\%$ & $8.0\%$ & & \\
$\Delta D_{s}$              & $11.0\%$ & $8.0\%$ & & \\
\hline
\end{tabular}
\end{table}

From 44 data points we obtain a photometric period of 1.96 days, which is exactly half the period reported by \cite{2004AcA....54..153P}. To check this discrepancy, we performed a re-analysis of the light curves available from the ASAS photometric catalog\footnote{http://www.astrouw.edu.pl/asas/} database. Firstly, we considered whether the difference maybe caused by the selection of the ASAS photometric data points, which are identified with the letters A to D according to the quality of the data. Using only the high-quality data available (marked with A) the PDM analysis yields a period $P = 1.96$ days which is exactly the same value obtained from our light curves. Second, we considered whether the discrepancy may be caused by the different algorithms used to calculate the period. However, using $\sim 50$ light curves obtained from the ASAS catalogue and performing a PDM analysis yields 100\% agreement with the values reported by \cite{2004AcA....54..153P}. Therefore we assume that
the entry in Table 2 from \cite{2004AcA....54..153P} must be erroneous.

Fig.~\ref{CPD518946} shown the light curves together with the best-fitting model solution corresponding to Sloan $r$ and $i$ filters. In addition, this figure shows the light curve obtained from the ASAS III catalogue with a period of 1.96 days. The light curves in both filters have similar amplitude variations of $\sim 16\%$. The primary and secondary minimum have different depths $\Delta D_{p} = 16\%$ and $\Delta D_{s} = 10\%$, which can be interpreted through a difference between the temperatures of the
components. Both partial eclipses have the same duration (0.214 in phase) and their minima are separated symmetrically (0.5 in phase). In consequence, we estimated the parameters of the system by assuming a circular orbit and a synchronous rotation. The primary component fills up its Roche lobe at about 95\% while the secondary at about 99\%, favouring a semi-detached configuration with an orbital inclination $i = 58.4^{\circ}$.

We fixed the effective temperature of the primary star $T_1 = 34550$\,K according to the color-index $B-V=-0.32$ obtained by \cite{2001A&A...369..527V}\footnote{Note that the color-index reported by van Houten was obtained in the Walraven photometric system. While the corresponding $V$ filter is similar to the Johnson $V$ passband in both peak and bandwith the $B$ filter has only half of the width of the Johnson $B$ passband.} corresponding to the spectral type O8.5\,V, reported originally by \cite{1981ARA&A..19..295B} and re-analyzed by \cite{1984ApJ...284..565H}; this spectral designation is also consistent with recent investigations by \cite{2005A&A...436.1049M}.

\begin{figure}
  \centering
    \includegraphics[angle=270,width=8cm]{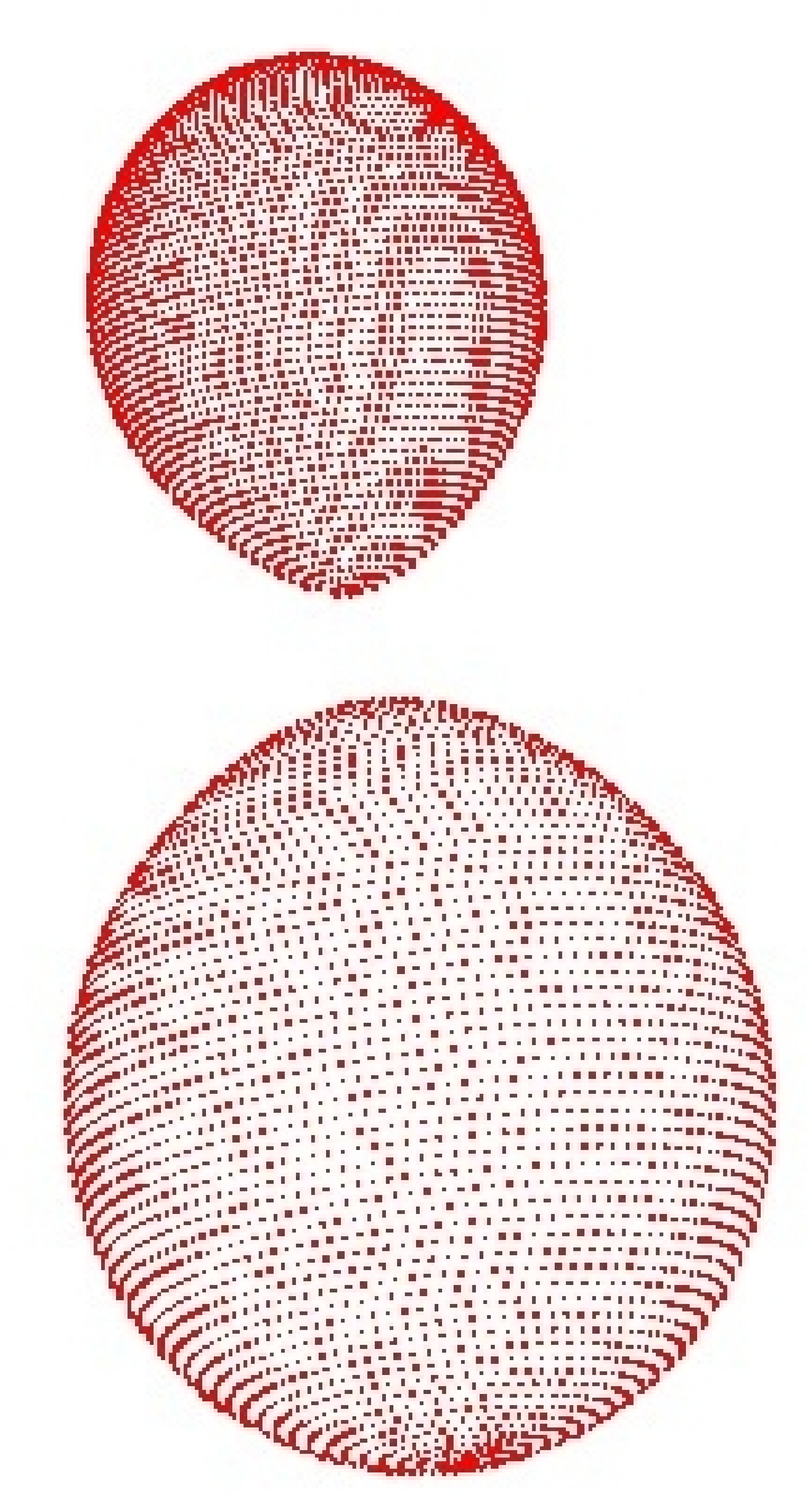}
  \caption{The graphical 3D view of the
  system CPD\,$-$\,51\degr\,8946 when the secondary star fills up its Roche lobe (bottom) was obtained
  with the parameters estimated from the light curve analysis.
  }
  \label{ASASV}
\end{figure}

The best fit to the light curve yields a mass ratio $q = M_1/M_2 = 2.8$ with a temperature of the secondary component $T_2 = 21500$\,K corresponding to a spectral type B2\,V (\citealt{1981ARA&A..19..295B}; \citealt{2010A&A...515A..74L}; \citealt{2012A&A...539A.143N}). As displayed in Fig.~\ref{CPD518946}, the 3D view Roche geometry of the system ($\phi = 0.63$) shows a transitional evolutionary state of the system for which both stars are close to filling their Roche Lobes. Most likely the future mass transfer will create a new contact binary system. A summary of the relevant parameters and the best-fitting values are listed in Table~\ref{table3}.

\subsection{{\bf HD\,319702}}

Located at a distance of 1.75\,kpc (\citealt{2010A&A...515A..55R}; \citealt{2012A&A...538A.142R}), HD\,319702 is a member of the cluster NGC\,6334. This region is considered as one of the most active regions of high-mass star formation in our Galaxy (\citealt{2009RMxAC..35...78T}; \citealt{2012A&A...538A.142R}). The area contains evolved optical but also embedded compact H\,{\sc ii} regions. HD\,319702 has been classified originally as a B1\,Ib by \cite{1978A&A....69...51N}; later measurements converted the spectral type into O8\,III(f) (\citealt{1982AJ.....87.1300W}; \citealt{2004ApJS..151..103M}; \citealt{2010A&A...521A..26P}). So far there was no hint for an eclipsing orbit.

\begin{figure}
  \centering
  \includegraphics[angle=0,width=\columnwidth]{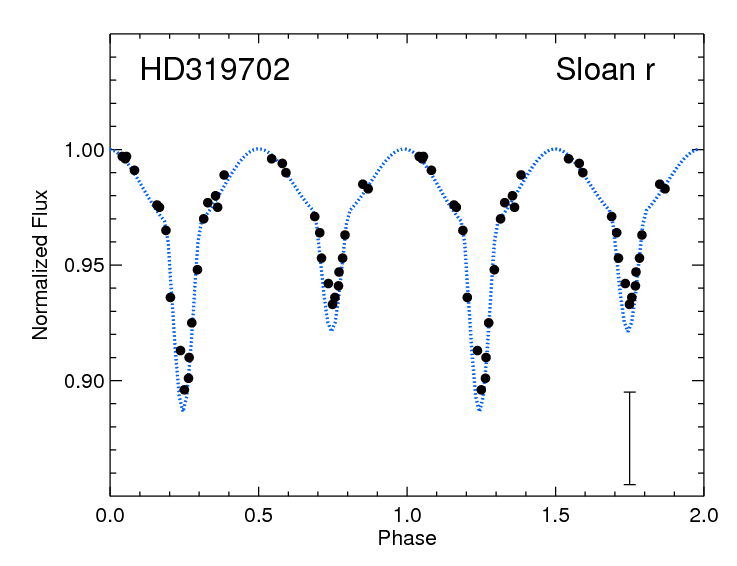}
  \includegraphics[angle=0,width=\columnwidth]{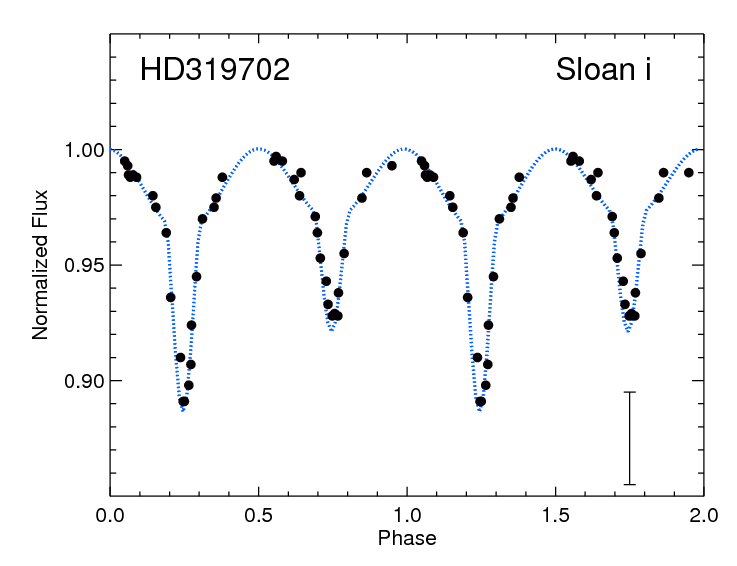}
 \includegraphics[angle=270,width=6cm]{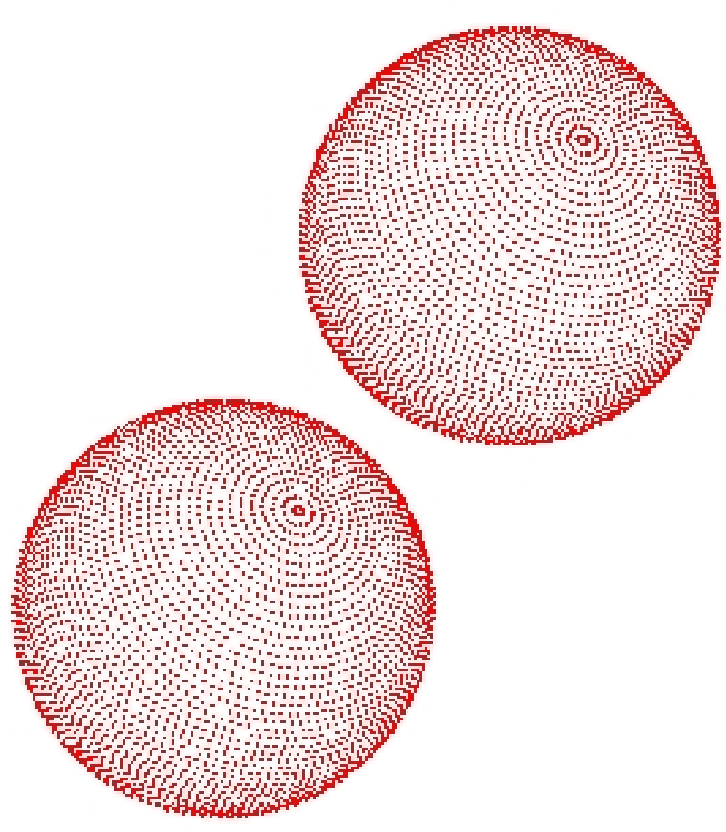}
  \caption{Observed (dots) and simulated (solid line) light curves in the Sloan~$r$ (top) and Sloan~$i$ (middle) bands, folded onto the orbital period ($2.01$ days) obtained from the PDM analysis. The error bar in the bottom of the plot represents the average measurement uncertainty for the data set. The graphical 3D view of the new detached binary system HD\,319702 at an orbital phase 0.5 is displayed at the bottom.}
  \label{HD319702}
\end{figure}

As displayed in Fig.~\ref{HD319702}, the light curves constructed from 38 epochs show nearly identical amplitude variations with primary and secondary minimum eclipses at different depths ($\Delta D_{p} = 11\%$, $\Delta D_{s} = 8\%$). This behaviour is characteristic for an eclipsing binary with different effective temperatures; the primary minimum is due to the eclipse of the
more luminous star by the less luminous companion. The eclipses are partial and symmetric with the same duration of 0.189 in phase. The primary and secondary minimum are separated symmetrically by 0.5 in phase suggesting a circular orbit.

Our PDM analysis yields a period of $P = 2.01$ days. Again, we have assumed a zero eccentricity and synchronous rotation for which the model provides the best fit to the system. Based on the spectral type determination of the primary star we adopted a value of $T_1 = 34000$\,K considering the observational $T_{eff}$ calibration by \cite{2005A&A...436.1049M} (his Table 5). We computed a set of orbital solutions where the best-fitting model favours an orbital inclination of 67.5$^{\circ}$ with a mass ratio of $q = M_1/M_2 = 1.0$. The system shows well-defined eclipses favouring a detached configuration, where both components fill up their Roche lobes at about 71\%. The temperature of the secondary component is $T_2 = 25200$\,K suggesting that this new system is most likely composed of an O8\,III + a B0.5\,V star if the calibration by \cite{1984ApJ...284..565H} is used for the secondary component. Considering the calibration reported by \cite{1981ARA&A..19..295B} and \cite{1988BAICz..39..329H} the system would turn into O8\,III + a B1\,V. A summary of the relevant parameters and the best-fitting values are listed in Table~\ref{table4}.

\section{Summary and conclusions}
\label{section_conclusions}

We have presented optical photometric observations for three high-mass eclipsing binaries performed during a six months monitoring campaign. A detailed analysis and modeling of the light curves was carried out within the framework of the Roche geometry. We find that all objects show light curve variations that occur on time-scales of less than 3 days and that the systems are well described by circular orbits (zero or negligible eccentricity). The individual results are:

\begin{itemize}

\item \object{Pismis\,24-1} is a detached system with an orbital inclination of $61.8^{\circ}$ and an orbital period of 2.36 days. We can confirm that the eclipsing binary system of Pismis\,24-1 is composed by at least one O+O pair.

\item \object{CPD\,$-$\,51\degr\,8946} is a semi-detached system with an orbital inclination of $58.4^{\circ}$ and an orbital period of 1.96 days. This system is most likely composed of an O8.5\,V and a B2\,V star and develops toward a contact system.

\item \object{HD\,319702} is a new high-mass eclipsing binary system with an detached configuration; its orbital inclination is $67.5^{\circ}$ and the period is 2.01 days. The components are well described by the association of an O8\,III and a  B1\,V star.

\end{itemize}

The current study has demonstrated that our photometric survey has the capability to detect O-type eclipsing binaries along the galactic plane in the brightness range $10 < R,I < 15$; more EB candidates will be presented in the future. Follow-up spectroscopic $RV$ studies of this sample are essential to determine the absolute parameters and to track the evolutionary state of individual systems. On the other hand, for the complete sample from the Galactic O-Star Catalogue ($V < 8$) we are currently monitoring all SB2 candidates (\citealt{2012MNRAS.424.1925C}) to detect further O-type EBs among this sample and to obtain precise light curves (Barr Dom\'{i}nguez et al., in prep.). From the few results available so far -- both in the literature and from our current study-- it seems that the important mass ratio parameter $q \sim 1$. However, there might be an observational bias since O stars have very high luminosities which prevent fainter companions from being detected. So far there are only four known high-mass binaries where $q > 2$ with the exceptional maximum of $q \sim 5.8$:
HD\,37022 (O5), HD\,53975 (B7\,Iab), HD\,199579 (O6\,V), and HD\,165246 (O8\,V) (see \citealt{2013A&A...550A...2M} and references therein).

This strongly corroborates the view that high-mass binaries are generally created during the star formation process and are not  a result of tidal capture. We expect that our study will increase the number of known O-type EBs substantially and that we can obtain a better census of the range of   $q = M_1 / M_2$ in the high-mass regime. Likewise it will be interesting to see at which stellar primary mass $q$ will significantly deviate from unity. For this reason we will extend our studies in the future also toward B-type binaries.

\begin{acknowledgements}

  This publication is supported as a project of the Nordrhein-Westf\"alische Akademie der Wissenschaften und der K\"unste in the framework of the academy program by the Federal Republic of Germany and the state Nordrhein-Westfalen.

  The observations at Cerro Armazones benefitted from the continuous support of the Universidad Cat\'{o}lica del Norte and from the care of the guardians Hector Labra, Gerardo Pino, Roberto Munoz, and Francisco Arraya.

  This research has made use of the NASA/IPAC Extragalactic Database (NED) which is operated by the Jet Propulsion Laboratory, California Institute of Technology, under contract with the National Aeronautics and Space Administration. This research has made use of the SIMBAD database, operated at CDS, Strasbourg, France. We thank the anonymous referee for his contructive comments and careful review of the manuscript.

\end{acknowledgements}

\bibliographystyle{aa}
\bibliography{high_mass}

\end{document}